\documentclass[useAMS,usenatbib]{mn2e}

\title[Excitation of stellar oscillations by GWs]{Excitation of non-radial stellar oscillations by gravitational waves: a first model}
\author[D. M. Siegel and M. Roth]{D. M. Siegel$^{1}$\thanks{E-mail:
daniel.siegel@kis.uni-freiburg.de; markus.roth@kis.uni-freiburg.de} and M. Roth$^{1}$\footnotemark[1]\\
$^{1}$Kiepenheuer-Institut f\"ur Sonnenphysik, Sch\"oneckstr. 6, 79104 Freiburg, Germany}
\begin{document}

\date{Accepted XXXX Received XXXX}

\pagerange{\pageref{firstpage}--\pageref{lastpage}} \pubyear{XXXX}

\maketitle

\label{firstpage}

\begin{abstract}
The excitation of solar and solar-like g modes in non-relativistic stars by arbitrary external gravitational wave fields is studied starting from the full field equations of general relativity. We develop a formalism that yields the mean-square amplitudes and surface velocities of global normal modes excited in such a way. The isotropic elastic sphere model of a star is adopted to demonstrate this formalism and for calculative simplicity. It is shown that gravitational waves solely couple to quadrupolar spheroidal eigenmodes and that normal modes are only sensitive to the spherical component of the gravitational waves having the same azimuthal order. The mean-square amplitudes in case of stationary external gravitational waves are given by a simple expression, a product of a factor depending on the resonant properties of the star and the power spectral density of the gravitational waves' spherical accelerations. Both mean-square amplitudes and surface velocities show a characteristic $R^8$-dependence (effective $R^2$-dependence) on the radius of the star. This finding increases the relevance of this excitation mechanism in case of stars larger than the Sun.
\end{abstract}


\section{Introduction}

Solar-like oscillations in non-relativistic stars are thought to be stochastically excited by turbulent convection in the outer layers of the star. Theoretical investigations (\citealt{Goldreich1977}; \citealt{Balmforth1992}; \citealt*{Goldreich1994}; \citealt{Samadi2001a}) identified two major mechanisms by which oscillations are driven: the first source term is due to the turbulent Reynolds stress and is thus related to the Reynolds stress tensor; the second mechanism stems from the advection of Eulerian entropy fluctuations by turbulent motions (the entropy source term). A generalized formalism by \citet{Samadi2001a} for radial modes, which attempts at combining both mechanisms in a consistent forced wave equation, was further extended by \citet{Belkacem2008} to the case of non-radial modes. Theoretically computed excitation rates for radial solar p modes are in good agreement with observational data from the GOLF instrument on board the SOHO spacecraft (\citealt{Samadi2003a,Samadi2003b}; \citealt{Belkacem2006a,Belkacem2006b}). Furthermore, radial solar p-mode peak heights were shown to be reasonably well modelled (\citealt{Chaplin2005}; \citealt{Houdek2006}). However, in case of solar g modes the entropy source term is negligible \citep{Belkacem2009} and quantitative estimates of mode amplitudes differ from each other by orders of magnitude, depending mainly on the assumed eddy-time correlation function, that is on how the turbulent eddies are time-correlated (\citealt*{Kumar1996}; \citealt{Belkacem2009}; see in particular the discussion in \citealt{Appourchaux2010}). Since the detection of solar g modes still remains controversial, theoretical predictions of their amplitudes are of utmost importance.

In the present paper, we consider an independent driving mechanism which we expect to be relevant for solar and solar-like g modes, especially in case of stars larger than the Sun that are nearby strong gravitational wave sources: excitation by external gravitational waves. Due to damping rates, $l=1$ and $l=2$ solar g modes are the most probable candidates for detection \citep{Belkacem2009}. As is shown below, this mechanism is able to excite quadrupolar eigenmodes ($l=2$) and thus it may influence the future detection of solar and solar-like g modes.

The excitation of normal modes by gravitational waves has already been studied for relativistic stars. These stars were shown to have a family of normal modes that are directly associated with the curvature of space--time, called gravitational-wave modes (w modes), and that have no Newtonian analogue in non-relativistic stars \citep{KokkotasSchutz1992}. \citet{Andersson1996} first showed that these normal modes can be excited by gravitational waves impinging on the star. This initial work was extended and refined by many other studies (see e.g. the review by \citealt{KokkotasSchmidt1999}, \citealt*{Bernuzzi2008} and references cited therein). Moreover, since the mid-1990s asteroseismology and gravitational wave physics were combined in order to gain information about the interior of relativistic stars (see e.g. \citealt{Andersson1998}; \citealt*{Benhar2004}). Even in case of non-relativistic stars such as the Sun, gravitational wave signals generated by g modes have been studied hoping that a future gravitational wave detector such as the Laser Interferometer Space Antenna (LISA) will be able to unequivocally detect solar g modes \citep*{Polnarev2009}.

However, a thorough theoretical investigation of an excitation mechanism for normal modes of non-relativistic stars due to gravitational waves has not been presented so far and is therefore worth studying even for its own theoretical interest. \citet{KhosroshahiSobouti} already considered this excitation mechanism, but concentrated on calculating energy absorption cross-sections. Other authors studied such a driving mechanism focusing on the excitation of elastic waves in solid planets or on the excitation of normal modes of resonant spheres used in gravitational wave detectors. In the present paper, however, we take a different perspective: motivated by the aim of calculating surface velocities for stars, which are interesting quantities for asteroseismology, we present a new way of solving this excitation problem.

The general idea of the present paper is to combine the physics of spherical resonant mass gravitational wave detectors (see e.g. \citealt{AshbyDreitlein}; \citealt{Lobo}; \citealt{Mag}, and references cited therein) with asteroseismology, i.e. we consider a star as a giant gravitational wave detector. The equation of motion for the internal velocity field of a star under the action of arbitrary external gravitational waves is derived and solved analytically. This analytical solution to the equation of motion enables us then to derive analytical expressions for the surface velocities, which in turn are crucial for assessing the astrophysical and asteroseismic relevance of the proposed excitation mechanism. However, we do not conduct a hydrodynamical calculation, but instead we take an isotropic elastic sphere as a very simple and crude model of a star for calculative simplicity. This resonant sphere model is ideally suited to demonstrate the formalism of the proposed excitation mechanism and it eventually yields the same analytical results as in a hydrodynamical computation. This is due to the fact that there is a similar equation of motion in both cases and it is also due to the spherical symmetry of the problem, since, as it turns out, the separation of the star's spheroidal normal modes into a horizontal and a radial component together with the orthogonality relation of the normal modes already constitutes the relevant information about the normal modes that is needed to analytically solve the equation of motion for the internal velocity field. Employing the elastic sphere model also makes this formalism applicable to the excitation of elastic waves in solid planets, which, however, is not the focus of the present paper.

The paper is organized as follows. In Section 2 we briefly discuss the adopted approximation of linearized gravity and some properties of gravitational waves in transverse-traceless gauge that are needed for the subsequent analysis. Following \citet{AshbyDreitlein}, we start from Einstein's equations of general relativity and the Bianchi identities in order to derive the inhomogeneous equation of motion for the internal velocity field of the star under the action of arbitrary external gravitational waves in Section 3. In this section we also discuss in detail the various assumptions and approximations that enter the computation. Section 4 is devoted to a very brief discussion of the normal modes, which are the solutions to the homogeneous equation of motion without damping and forcing terms. In Section 5 we analytically solve the inhomogeneous equation of motion and calculate the square amplitudes of the excited normal modes. These square amplitudes finally allow for an analytical computation of surface velocities in Section 6. Conclusions and prospects are presented in Section 7; an appendix is added to shorten the calculation in Section 5.

\section{Gravitational Waves}

For the present analysis we assume that the linearized theory of general relativity is applicable, so that the metric takes the form
\[
	g_{\mu\nu}=\eta_{\mu\nu}+h_{\mu\nu},
\]
where $\vert h_{\mu\nu}\vert \ll 1$ and $\eta_{\mu\nu}=\rm{diag}(-1,+1,+1,+1)$. As is shown in Section 3, we are free to choose the transverse-traceless (TT) gauge, defined by\footnote{Throughout this paper, Greek indices take space--time values 0,1,2,3, whereas Latin indices take spatial values 1,2,3 only. Repeated indices are summed over.}
\begin{equation}
		h^{0\mu}=0,\;\;\;h^i_{\phantom{i}i}=0,\;\;\;\partial^jh_{ij}=0. \label{eq:tt_gauge}
\end{equation}
The gravitational waves incident on the star under consideration satisfy the linearized field equations in vacuo, which in TT gauge read
\[
	\sq h_{ij}=0.
\]

From equation (\ref{eq:tt_gauge}), it is obvious that $h_{ij}$ is a spin-two tensor field and thus in an expansion in spherical harmonics, there are solely contributions from $l=2$:
\begin{equation}
	h_{ij}=\sum_{m=-2}^{2}h_m\mathcal{Y}_{ij}^{2m}, \label{eq:expan.sph.harm.}
\end{equation}
where $h_m$ are the five independent spherical components of $h_{ij}$, and $\mathcal{Y}_{ij}^{2m}$ are components of the following matrices, which form a basis for the five-dimensional space of traceless symmetric tensors:
\begin{eqnarray}
	\mathcal{Y}_{ij}^{2,\pm 2}&=&\sqrt{\frac{15}{32\pi}}\left( \begin{array}{ccc} 
	1 & \pm \rmn{i} & 0 \\ 
	\pm \rmn{i} & -1 & 0 \\
	0 & 0 & 0
	\end{array} \right)_{ij},\nonumber\\
	\mathcal{Y}_{ij}^{2,\pm 1}&=&\mp\sqrt{\frac{15}{32\pi}}\left( \begin{array}{ccc} 
	0 & 0 & 1 \\ 
	0 & 0 & \pm \rmn{i} \\
	1 & \pm \rmn{i} & 0
	\end{array} \right)_{ij},\nonumber\\
	\mathcal{Y}_{ij}^{20}&=&\sqrt{\frac{5}{16\pi}}\left( \begin{array}{ccc} 
	-1 & 0 & 0 \\ 
	0 & -1 & 0 \\
	0 & 0 & 2
	\end{array} \right)_{ij}.\nonumber	
\end{eqnarray}
These matrices are linked to spherical harmonics via
\begin{equation}
	Y_{2m}(\theta,\phi)=\mathcal{Y}_{ij}^{2m}n_in_j, \label{eq:sph.harm}
\end{equation} 
where $n_i$ are the components of the radial unit vector $\bmath{n}=(\sin\theta\cos\phi,\sin\theta\sin\phi,\cos\theta)$, and the orthogonality relation is given by
\begin{equation}
	\sum_{ij}\mathcal{Y}_{ij}^{2m}\left(\mathcal{Y}_{ij}^{2m'}\right)^*=\frac{15}{8\pi}\delta^{mm'} \label{eq:orth.rel}
\end{equation}
(see, e.g., \citealt{Mag}).

For later use we mention two additional important relations. Multiplying equation (\ref{eq:expan.sph.harm.}) by $n_in_j$, summing over $i$ and $j$ and inserting equation (\ref{eq:sph.harm}), we obtain
\begin{equation}
	h_{ij}n_in_j=\sum_{m=-2}^{2}h_mY_{2m}(\theta,\phi), \label{eq:rel_1}
\end{equation}
and inverting equation (\ref{eq:sph.harm}) with the help of equation (\ref{eq:orth.rel}) yields
\begin{equation}
	n_in_j-\frac{1}{3}\delta_{ij}=\sum_{m=-2}^{2}c_{ij}^m\, Y_{2m}(\theta,\phi), \label{eq:rel_2}
\end{equation}
where the factor of 1/3 is fixed by the requirement that the left hand side be traceless, since the coefficients $c_{ij}^m$ are given by $c_{ij}^m=\frac{8\pi}{15}(\mathcal{Y}_{ij}^{2m})^*$.

\section{Equation of motion}

In order to derive the equation of motion for the internal velocity field of a star under the influence of external gravitational waves, we start with the full field equations of general relativity and the Bianchi identities, which together imply the conservation equations of energy and momentum,
\begin{equation}
	T^{\mu\nu}_{\phantom{\mu\nu};\nu}=0, \label{eq:EMC}
\end{equation}
where $T^{\mu\nu}$ are the components of the stress-energy tensor of the star under consideration. The equation of motion is obtained from the spatial components of equation (\ref{eq:EMC}):
\begin{equation}
	\frac{\partial T^{0i}}{\partial x^0}+\frac{\partial T^{ij}}{\partial x^j}=-\Gamma^i_{\mu\rho}T^{\mu\rho}-\Gamma^\mu_{\mu\rho}T^{\rho i}. \label{eq:EOM1}
\end{equation}

Since the centre of mass of the star will move on a geodesic in space--time, it proves useful to work in Fermi normal coordinates with the origin at the centre of mass at all times. In this reference frame
\[
	\Gamma^\rho_{\mu\nu}[P(t)]=0, \;\;\; \partial_0\Gamma^\rho_{\mu\nu}[P(t)]=0,
\]
where $P(t)$ denotes the centre of mass at time $t$. Using these expressions one finds for the Riemann tensor
\[
	R^i_{\phantom{i}0j0}[P(t)]=\partial_j\Gamma^i_{00}[P(t)].
\]

Alternatively, within the framework of the linearized theory -- which we assume is valid here -- the Riemann tensor is invariant, rather than just covariant, and thus can be evaluated in any preferred frame. Consequently, choosing the TT frame for convenience one has from the linearized theory the following expression for the components $R^i_{\phantom{i}0j0}$ of the Riemann tensor in terms of the metric:
\[
	R^i_{\phantom{i}0j0}=R_{i0j0}=-\frac{1}{2c^2}\ddot{h}_{ij},
\]
where $h_{ij}$ are the gravitational wave components of the metric in TT gauge.

It is now assumed that the diameter $d$ of the star is much smaller than the typical length scale $\lambda/2\pi$ over which the incident gravitational radiation changes substantially. Clearly, this is an assumption that in practice has to be checked case by case. Due to the fact that under this assumption the components $h_{ij}$ have essentially no spatial dependence over the volume of the star, we obtain the following relation
\[
	\partial_j\Gamma^i_{00}=-\frac{1}{2c^2}\ddot{h}_{ij},
\]
which can be integrated to give
\[
	\Gamma^i_{00}=-\frac{1}{2c^2}\ddot{h}_{ij}x^j.
\]

Moreover, we assume that the internal motions of the star are non-relativistic. In this Newtonian approximation, only $T^{00}$ terms need be retained on the right-hand side of equation (\ref{eq:EOM1}) and $T^{00}$ is given by $T^{00}=\rho c^2$, where $\rho$ is the equilibrium proper mass density of the star. Identifying $T^{0i}/c$ as the non-relativistic momentum density given by $\rho \bmath{v}$, and $T^{ij}$ as the negative non-relativistic stress tensor, $T^{ij}=-\sigma^{ij}$, we arrive at
\begin{equation}
	\rho\frac{\partial v_i}{\partial t}=\frac{\partial\sigma_{ij}}{\partial x^j}+\frac{1}{2}\rho\ddot{h}_{ij}x^j, \label{eq:EOM2}
\end{equation}
where $\bmath{v}$ with components $v_i$ denotes the internal velocity field of the star.

For the reasons stated above in Section 1, the star is treated as an isotropic elastic sphere. This requires $\rho(\bmath{x})=\rho(r)$. Due to the external force exerted by the gravitational waves, an infinitesimal volume element of the elastic sphere centred at position $\bmath{x}$ will be displaced according to $\bmath{x}+\bmath{u}(\bmath{x},t)$, where we assume the displacements to be sufficiently small such that the linear theory of elasticity is appropriate. To this approximation and neglecting self-stresses caused by the intrinsic gravitational field (see also the discussion in Section 4), the elastic stress tensor for isotropic media is given by
\begin{equation}
	\sigma_{ij}=\left(\lambda+\lambda'\frac{\partial}{\partial t}\right)u_{kk}\delta_{ij}+2\left(\mu+\mu'\frac{\partial}{\partial t}\right)u_{ij}, \label{eq:stress_ten}
\end{equation}   
where $u_{lm}\equiv (1/2)(\partial_l u_m+\partial_m u_l)$ and $\lambda$ and $\mu$ are the usual Lam\'e coefficients \citep{LanLif}. The positive constants $\lambda'$ and $\mu'$ parametrize the viscous properties of the elastic sphere. Note that equation (\ref{eq:stress_ten}) only holds in Fermi normal coordinates.

Since the velocity field of a star is the easier measurable quantity than the displacements themselves, e.g. by Doppler techniques, we differentiate equation (\ref{eq:EOM2}) with respect to time in order to obtain the equation of motion in terms of the velocity field:
\begin{equation}
	\rho \left(\frac{\partial^2}{\partial t^2}-\mathcal{L}\right)\bmath{v}+\mathcal{D}(\dot{\bmath{v}})=\frac{\partial{\bmath{f}}}{\partial t}, \label{eq:EOM3}
\end{equation} 
where $\bmath{f}(\bmath{x},t)$, with components given by 
\begin{equation}
	f_i=\frac{1}{2}\rho\ddot{h}_{ij}x^j, \label{eq:drivingterm}
\end{equation}
denotes the external driving force per unit volume exerted by the gravitational waves. The linear differential operators $\mathcal{L}$ and $\mathcal{D}$ act on arbitrary differentiable vector fields $\bmath{X}$ according to
\begin{eqnarray}
	\mathcal{L}(\bmath{X})&=&\frac{1}{\rho}\left[(\lambda +\mu)\nabla(\bmath{\nabla\cdot X})+\mu\nabla^2\bmath{X}\right]\nonumber\\
	\mathcal{D}(\bmath{X})&=&-(\lambda' +\mu')\nabla(\bmath{\nabla\cdot X})-\mu'\nabla^2\bmath{X}. \nonumber
\end{eqnarray}

The model of viscosity adopted here corresponds to the Kelvin--Voigt model considered by \citet{Ortega} and \citet{LoboOrtega}. For the present paper, however, it suffices to consider the damping operator $\mathcal{D}$ as some general abstract operator that is linear in its argument and does not involve any time derivatives. These are the only properties that enter and affect the calculation conducted in Section 5. Hence, in principle, any other (phenomenological) model may be employed, provided that the resulting damping operator has the aforementioned properties.

\section{Normal modes of an elastic sphere}

The normal modes of an elastic sphere form a basis of solutions to the homogeneous equation of motion without a damping term,
\begin{equation}
	\left(\frac{\partial^2}{\partial t^2}-\mathcal{L}\right)\bmath{u}=0. \label{eq:EOM_hom}
\end{equation}
This equation of motion has to be supplemented with the appropriate boundary condition
\begin{equation}
	\sigma_{ij}n_j=0 \label{eq:boundary}
\end{equation}
at $r=R$, where $\lambda'$ and $\mu'$ are set to zero, $n_i\equiv x_i/r$ is the unit normal in Cartesian coordinates and $R$ denotes the radius of the sphere. In order to find the normal modes, one can assume the displacement to have a harmonical time dependence,
\begin{equation}
	\bmath{u}(\bmath{x},t)=\mathbf{\Psi}(\bmath{x})\rmn{e}^{-\rmn{i}\omega t}. \label{eq:sepAn}
\end{equation}
Inserting equation (\ref{eq:sepAn}) into equation (\ref{eq:EOM_hom}) yields the eigenvalue problem
\begin{equation}
	\mathcal{L}\mathbf{\Psi}(\bmath{x})=-\omega^2\mathbf{\Psi}(\bmath{x}). \label{eq:eigenvalprob}
\end{equation}
Solving for the eigenfunctions and eigenfrequencies of $\mathcal{L}$ subject to the boundary conditions given by equation (\ref{eq:boundary}) is a standard problem in elasticity theory und thus shall not be reconsidered here. We solely quote the final result below for later use in Section 5 (cf. e.g. \citealt{AshbyDreitlein}; \citealt{Lobo}; \citealt{Mag}). 

There are two families of solutions to equation (\ref{eq:eigenvalprob}). The spheroidal modes can be written as
\begin{equation}
	\mathbf{\Psi}^{S}_{nlm}(r,\Theta,\phi)=a_{nl}(r)Y_{lm}\bmath{e}_r-b_{nl}(r)\rmn{i}\bmath{e}_r\bmath{\times L}Y_{lm}, \label{eq:sphmodes}
\end{equation}
where $\bmath{L}\equiv-\rmn{i}\bmath{x\times\nabla}$ is the angular momentum operator. The normalization is fixed by requiring
\begin{equation}
	\int_{V} \rmn{d}^3x\, \rho (\mathbf{\Psi}^{S}_{nlm})^*\bmath{\cdot}\mathbf{\Psi}^S_{nlm}=I, \label{eq:normalization}
\end{equation}  
where $I$ is normally set equal to the mass $M$ of the sphere and $V$ denotes its volume.
For $l=0$, the spheroidal modes are purely radial, that is, $b_{n0}(r)=0$. The toroidal modes are given by the expression
\begin{equation}
	\mathbf{\Psi}^{T}_{nlm}(r,\Theta,\phi)=c_{nl}(r)\rmn{i}\bmath{L}Y_{lm}, \label{eq:tor_modes}
\end{equation} 
where $l\ge 1$, and are normalized in the same way as the spheroidal modes. The orthogonality and normalization properties of the eigenfunctions of $\mathcal{L}$ can be summarized as
\begin{equation}
	\int_{V} \rmn{d}^3x\, \rho \mathbf{\Psi}_N^*\bmath{\cdot}\mathbf{\Psi}_{N'}=I\delta_{NN'}, \label{eq:normalization2}
\end{equation}
where $N\equiv\{nlm;S\,\rmn{or}\,T\}$ is an abridged index. It is important to stress that in the present paper we are not interested in detailed analytical expressions for the coefficient functions $a_{nl}(r)$, $b_{nl}(r)$ and $c_{nl}(r)$, we rather treat them as variables, i.e. we express the final results as functions of these coefficient functions.

In deriving equation (\ref{eq:stress_ten}), effects of self-stress caused by the intrinsic gravitational field of the sphere were neglected. However, these effects are expected to be non-negligible for spheres as massive as astronomical objects \citep{AshbyDreitlein}. Including these effects into the present formalism yields a different differential operator $\mathcal{L}$ in equation (\ref{eq:EOM3}),
\begin{eqnarray}
	\tilde{\mathcal{L}}(\bmath{X})&=&\frac{1}{\rho_0}\left\{ (\lambda +\mu)\nabla(\bmath{\nabla\cdot X})+\mu\nabla^2\bmath{X}\right.\nonumber\\
	&&+\left[(\bmath{X\cdot\nabla})\rho_0+\rho_0(\bmath{\nabla\cdot X})\right]\nabla V_0 \nonumber\\
	&&- \left.\rho_0\nabla V^{(1)}-\nabla\left[\rho_0(\bmath{X\cdot\nabla})V_0\right]\right\}\nonumber
\end{eqnarray}
\citep{AshbyDreitlein}, where uniform elastic constants were assumed and all quantities $Q$ such as pressure, density, gravitational potential $V$, etc. have been expanded to first order in $h_{ij}$, $Q=Q_0+Q^{(1)}$, where $Q_0$ denotes the zeroth-order quantity and $Q^{(1)}$ the first-order correction. Due to spherical symmetry, however, the eigenfunctions of $\tilde{\mathcal{L}}$ are of the same type as in equation (\ref{eq:sphmodes}) and (\ref{eq:tor_modes}), but having different coefficient functions $\tilde{a}_{nl}(r)$, $\tilde{b}_{nl}(r)$ and $\tilde{c}_{nl}(r)$. In case of uniform density, $\rho_0=\rmn{const.}$, the spheroidal coefficient functions $\tilde{a}_{nl}(r)$ and $\tilde{b}_{nl}(r)$ are computed explicitly by \citet{AshbyDreitlein}. Since details of the analytical expressions of $a_{nl}(r)$, $b_{nl}(r)$ and $c_{nl}(r)$ do not enter and affect the calculations in Section 5 and 6, we simply assume eigenfunctions of the type given by equation (\ref{eq:sphmodes}) and (\ref{eq:tor_modes}), regardless of whether effects of self-stress have been included or not.

\section{Internal Velocity Field}

In this section, the (complex) internal velocity field $\bmath{v}(\bmath{x},t)$ of the star due to the external gravitational waves is calculated from equation (\ref{eq:EOM3}). Since the eigenfunctions of $\mathcal{L}$ form a complete orthogonal set, the complex velocity field can be expanded as
\begin{equation}
	\bmath{v}(\bmath{x},t)=\sum_{N}(-\rmn{i}\omega_N)A_N(t)\mathbf{\Psi}_N(\bmath{x})\rmn{e}^{-\rmn{i}\omega_N t}, \label{eq:velexpan}
\end{equation}
where $A_N$ denote complex time-dependent expansion coefficients -- the amplitudes of the velocity eigenmodes -- and $N\equiv\{nlm;S\,\rmn{or}\,T\}$ summarizes all indices of a particular eigenfunction. 

The individual amplitudes $A_N$ of the velocity eigenmodes are determined by solving equation (\ref{eq:EOM3}) for each velocity eigenmode 
\begin{equation}
	\bmath{v}_N(\bmath{x},t)=-\rmn{i}\omega_N A_N(t)\mathbf{\Psi}_N(\bmath{x})\rmn{e}^{-\rmn{i}\omega_N t} \label{eq:vel_eigenmode}
\end{equation} 
(cf. equations (\ref{eq:sepAn}) and (\ref{eq:velexpan})) separately. In the following, weak damping is assumed, $\eta_N\ll \omega_N$ for all $N$, where $\eta_N$ is the damping rate. In this limit one can make the approximation $|\partial A_N/\partial t|\ll\omega_N|A_N|$. Thus, substituting equation (\ref{eq:vel_eigenmode}) into equation (\ref{eq:EOM3}) and applying this approximation yields:
\[
	-2\omega_N^2\rho\frac{\rmn{d}A_N}{\rmn{d}t}\rmn{e}^{-\rmn{i}\omega_N t}\mathbf{\Psi}_N -\omega_N^2 A_N \rmn{e}^{-\rmn{i}\omega_N t}\mathcal{D}(\mathbf{\Psi}_N)=\frac{\partial{\bmath{f}}}{\partial t}.
\]
The equation of motion for a particular eigenmode amplitude $A_N$ is now obtained by multiplying this equation by $\mathbf{\Psi}_N^*(\bmath{x})$, integrating over the stellar volume $V$ and using the normalization condition, equation (\ref{eq:normalization2}), to give
\begin{equation}
	\frac{\rmn{d}A_N}{\rmn{d}t}+\sigma_N A_N =\frac{-\rmn{e}^{\rmn{i}\omega_N t}}{2\omega_N^2I}\int_V\rmn{d}^3x\,\mathbf{\Psi}_N^*\bmath{\cdot}\frac{\partial{\bmath{f}}}{\partial t}.\label{eq:EOM_A}
\end{equation}
Here we have set
\[
	\sigma_N\equiv\frac{1}{2I}\int_V\rmn{d}^3x\,\mathbf{\Psi}_N^*\bmath{\cdot}\mathcal{D}(\mathbf{\Psi}_N)\equiv \eta_N +\rmn{i}\delta\omega_N,
\]
where $\eta_N$ and $\delta\omega_N$ denote the real and imaginary part, respectively. Introducing the damped eigenfrequency 
\[
	\bar{\omega}_N\equiv\omega_N+\delta\omega_N,
\]
the solution to equation (\ref{eq:EOM_A}) after integration by parts with respect to time is given by
\[
	A_N(t)=\frac{\rmn{i}\rmn{e}^{-\sigma_N t}}{2\omega_N I}\left(1+\frac{\delta\omega_N}{\omega_N}\right)\mskip-5mu\int_{-\infty}^{t}\mskip-20mu\rmn{d}t_1\mskip-5mu\int_V\mskip-5mu\rmn{d}^3x\,\rmn{e}^{(\eta_N +\rmn{i}\bar{\omega}_N) t_1}\mathbf{\Psi}_N^*\bmath{\cdot f},
\]
where we have assumed that the driving term $\bmath{f}(\bmath{x},t)$ is `turned on' at $t=-\infty$.
Hence, after inserting equation (\ref{eq:drivingterm}), one immediately finds
\begin{eqnarray}
	A_N(t)&\mskip-20mu=&\mskip-20mu\frac{\rmn{i}\rmn{e}^{-\sigma_N t}}{4\omega_N I}\left(1+\frac{\delta\omega_N}{\omega_N}\right)\int_{-\infty}^{t}\rmn{d}t_1\,\rmn{e}^{(\eta_N +\rmn{i}\bar{\omega}_N) t_1} \nonumber\\ &\times&\mskip-10mu\ddot{h}_{ij}(t_1)\int_V\rmn{d}^3x\,\rho(r)[\mathbf{\Psi}^*_N(\bmath{x})]^i x^j. \label{eq:A_1}
\end{eqnarray}
Note that thanks to the long-wavelength approximation knowledge of the gravitational wave field is not required for evaluating the spatial integral, it can rather be computed by using properties of the elastic sphere only. The explicit computation of the spatial integral and the contraction with the gravitational wave tensor is discussed in Appendix A. It is found that gravitational waves do solely excite spheroidal modes with $l=2$, that is $A_{nlm}^T(t)=0$, $A_{nlm}^S(t)=0$ for $l\ne2$ and
\begin{equation}
	A_{n2m}^S(t)=\frac{\rmn{i}R^4\chi_n}{4\omega_N I}\mathcal{F}\rmn{e}^{-\sigma_N t}\int_{-\infty}^{t}\rmn{d}t_1\,\rmn{e}^{(\eta_N +\rmn{i}\bar{\omega}_N) t_1} \ddot{h}_m(t_1). \label{eq:A_2}
\end{equation}
In order to shorten the notation, the abbreviation $\mathcal{F}\equiv(1+\delta\omega_N/\omega_N)$ was introduced and $N$ was set equal to $\{n2m;S\}$. Furthermore, $\ddot{h}_m$ denotes the spherical component of $\ddot{h}_{ij}$ corresponding to the given azimuthal order $m$ as defined in equation (\ref{eq:expan.sph.harm.}), $R$ denotes the stellar radius and $\chi_n$ is given by the following integral
\[
	\chi_n\equiv\int_{0}^{1}\rmn{d}z\,\rho(z)z^3\left[a_{n2}(z)+3b_{n2}(z)\right],
\]
where $z\equiv r/R$. Clearly, $\chi_n$ models the sphere's ability to resonate in a particular eigenmode due to the external gravitational waves. Note that from equation (\ref{eq:A_2}), a specific eigenmode is only sensitive to the component $h_m$ of the gravitational waves having the same azimuthal order $m$.

After substitution into equation (\ref{eq:velexpan}), these amplitudes of the velocity eigenfunctions given by equation (\ref{eq:A_2}) entirely determine the internal complex velocity field of the star under the action of an external gravitational wave field. However, the internal (complex) velocity field of a star is not a direct observable. As shown in Section 6, relevant quantities with regard to experiments are the square amplitudes $|A_N(t)|^2$ of the velocity eigenmodes, rather than the amplitudes themselves. Thus, for the remainder of this section we proceed by computing the square amplitudes and start with multiplying equation (\ref{eq:A_2}) by its complex conjugate to obtain
\begin{eqnarray}
	|A_{n2m}(t)|^2&\mskip-20mu=&\mskip-20mu\frac{R^8\chi_n^2}{16\omega_N^2 I^2}\mathcal{F}^2\mskip-5mu\int_{-\infty}^{t}\mskip-20mu\rmn{d}t_1\rmn{d}t_2\,\rmn{e}^{\eta_N(t_1+t_2-2t)}\rmn{e}^{-\rmn{i}\bar{\omega}_N(t_2-t_1)} \nonumber\\
	 &\times&\mskip-10mu\ddot{h}_m(t_1)\ddot{h}_m^*(t_2). \label{eq:A2_3}
\end{eqnarray}

By defining new variables,
\[
	t_0\equiv\frac{t_1+t_2}{2},\;\;\;\tau\equiv t_2-t_1,
\]
equation (\ref{eq:A2_3}) can be transformed into
\[
	|A_N(t)|^2\mskip-5mu=\mskip-5mu\frac{R^8\chi_n^2}{16\omega_N^2 I^2}\mathcal{F}^2\mskip-5mu\int_{-\infty}^{t}\mskip-20mu\rmn{d}t_0\,\rmn{e}^{2\eta_N(t_0-t)}\mskip-5mu\int_{-2(t-t_0)}^{2(t-t_0)}\mskip-40mu\rmn{d}\tau\,K_m^{t_0}(\tau)\rmn{e}^{-\rmn{i}\bar{\omega}_N \tau},
\]
where we have introduced the correlation function
\[
	K_m^{t_0}(\tau)\equiv\ddot{h}_m\left(t_0-\frac{\tau}{2}\right)\ddot{h}_m^*\left(t_0+\frac{\tau}{2}\right).
\]
It is now assumed that the gravitational wave field is stationary, i.e. the correlation function $K_m^{t_0}$ is independent of $t_0$, $K_m^{t_0}=K_m(\tau)$. Since we are not interested in the transient behaviour of the system after the driving force of the gravitational waves has been `turned on' at $t=-\infty$, but rather in the stationary late-time behaviour, we formally consider the limit $t\rightarrow \infty$. In this limit the integrals over $t_0$ and $\tau$ can be performed independently of each other to give the final result
\begin{eqnarray}
	|A_{n2m}^S|^2&\equiv& \lim_{t\rightarrow\infty}|A_{n2m}^S(t)|^2 \nonumber\\
	&=&\frac{R^8\chi_n^2}{32\eta_N\omega_N^2 I^2}\left(1+\frac{\delta\omega_N}{\omega_N}\right)^2 P_m(\bar{\omega}_N), \label{eq:A2_f1}
\end{eqnarray}
where $P_m(\omega)$ is the Fourier transform of the correlation function $K_m(\tau)$,
\[
	P_m(\omega)\equiv\int_{-\infty}^{\infty}\rmn{d}\tau\,K_m(\tau)\,\rmn{e}^{-\rmn{i}\omega\tau}.
\]
Due to the assumption of a stationary gravitational wave field, $K_m(\tau)$ is the auto-correlation function of the spherical acceleration $\ddot{h}_m(t)$ and thus, $P_m(\omega)$ is the power spectral density of $\ddot{h}_m(t)$.

Normally, in the case of weak damping the frequency shifts $\delta\omega_N$ of the eigenfrequencies due to damping are expected to be small, $\delta\omega_N\ll \omega_N$, in which case we have
\begin{equation}
	|A_{n2m}^S|^2=\frac{R^8\chi_n^2}{32\eta_{N}\omega_{N}^2 I^2}P_m(\omega_{N}), \label{eq:A2_f2}
\end{equation}
where, again, $N=\{n2m;S\}$. Note the very sensitive $R^8$-dependence on the radius of the star. However, when applying the normalization condition (\ref{eq:normalization}) this will be reduced to an effective $R^2$-dependence. Note in particular the remarkable simplicity of this result, equations (\ref{eq:A2_f1}) and (\ref{eq:A2_f2}). Apart from the aforementioned effective $R^2$-dependence and numerical pre-factors, the square amplitudes of the velocity eigenfunctions are simply obtained by multiplying the incoming power spectral density of the spherical accelerations corresponding to a given azimuthal order $m$ with the quantity $\chi^2$, which, as stated above, describes the resonance properties of the star for a given radial order $n$ in case of excitation by gravitational waves. The appearance of $P_m(\omega)$ in equation (\ref{eq:A2_f1}) and (\ref{eq:A2_f2}) can be interpreted as an explicit manifestation of the tidal force nature of the excitations.

\section{Surface velocity field}

As already stated in Section 5, the quantity accessible to experiments is the surface velocity field of a star, rather than its internal (complex) velocity field. 

We define the mean-square surface velocity $v_{N,s}^2$ for each mode $N=\{n2m;S\}$ that is excited by
\begin{equation}
	v_{N,s}^2\equiv\bigg\langle\frac{1}{2}\int\rmn{d}\Omega\,\bmath{v}_N(\bmath{x},t)\bmath{\cdot} \bmath{v}_N^*(\bmath{x},t)\bigg\rangle(R), \label{eq:def_surf_vel}
\end{equation}
where $\bmath{v}_N(\bmath{x},t)$ is the internal complex velocity field of the given velocity eigenmode, 
\begin{equation}
	\bmath{v}_N(\bmath{x},t)=-\rmn{i}\omega_N A_N(t)\mathbf{\Psi}_N(\bmath{x})\rmn{e}^{-\rmn{i}\omega_N t}, \label{eq:vel_eigenmode2}
\end{equation}
and $\langle\rangle$ denotes time average. Note that this definition is equivalent to the one employed by \citet{Belkacem2009}. Inserting equation (\ref{eq:vel_eigenmode2}) into the definition given in equation (\ref{eq:def_surf_vel}) yields
\begin{equation}
	v_{N,s}^2=\frac{1}{2}\omega_N^2 \big\langle|A_N(t)|^2\big\rangle \left[a_{n2}^2(R)+6b_{n2}^2(R)\right]. \label{eq:v_s_1}
\end{equation}
As shown in Section 5, for a stationary gravitational wave field the square amplitudes $|A_N(t)|^2$ become time independent in the late-time limit $t\rightarrow \infty$, in which case by setting
\[
	\Psi_N^2(r)\equiv a_{n2}^2(r)+6b_{n2}^2(r)
\]
and inserting equation (\ref{eq:A2_f1}) we obtain
\begin{equation}
	v_{N,s}^2=\frac{R^8\chi_n^2 P_m(\bar{\omega}_N)}{64\eta_N I^2}\left(1+\frac{\delta\omega_N}{\omega_N}\right)^2 \Psi_N^2(R). \label{eq:v_s_f1}
\end{equation} 
Again, for small frequency shifts, $\delta\omega_N\ll \omega_N$, this expression reads
\begin{equation}
	v_{N,s}^2=\frac{R^8\chi_n^2 P_m(\omega_N)}{64\eta_N I^2}\Psi_N^2(R). \label{eq:v_s_f2}
\end{equation}
Alternatively, expressing the mean-square surface velocity in terms of the power that goes into each mode,
\begin{eqnarray}
	P_N&\mskip-15mu=&\mskip-15mu 2\eta_N\bigg\langle\int_{0}^{M}\rmn{d}m\,\frac{1}{2}\bmath{v}_N(\bmath{x},t)\bmath{\cdot} \bmath{v}_N^*(\bmath{x},t)\bigg\rangle\nonumber\\
	&\mskip-15mu=&\mskip-15mu\eta_N\omega_N^2 I\big\langle|A_N(t)|^2\big\rangle, \label{eq:power_1}
\end{eqnarray}
one has from equation (\ref{eq:v_s_1}) the relation
\begin{equation}
	v_{N,s}^2=\frac{P_N}{2\eta_N I}\Psi_N^2(R), \label{eq:v_s_power}
\end{equation}
where in the late-time limit, for small frequency shifts and hence with the help of equations (\ref{eq:A2_f2}) and (\ref{eq:power_1}) the power is explicitly given by
\begin{equation}
	P_N=\frac{R^8\chi_n^2}{32 I}P_m(\omega_N).\label{eq:power}
\end{equation}
Note again the characteristic $R^8$-dependence (effective $R^2$-dependence) on the radius of the star present in the expressions for the power $P_N$ and the surface velocity $v_{N,s}^2$. The important relations derived in this section, equations (\ref{eq:v_s_f1}), (\ref{eq:v_s_f2}), (\ref{eq:v_s_power}) and (\ref{eq:power}), are entirely analogous to the expressions obtained by \citet{Samadi2001a}, \citet*{Samadi2001b} and \citet{Belkacem2008,Belkacem2009} in case of stochastic excitation by turbulent convection.

\section{Conclusions}

In the present paper, an excitation mechanism has been established that yields the oscillation amplitudes of stellar oscillation modes when they are excited by arbitrary external gravitational waves. This mechanism is expected to be relevant for solar and solar-like g modes, especially in case of stars larger than the Sun that are nearby strong gravitational wave sources.

Starting from the general relativistic field equations, the inhomogeneous equation of motion for the internal (complex) velocity field of a non-relativistic star under the action of arbitrary external gravitational waves was derived. In this derivation, several assumptions and approximations have been made. It was assumed that the linearized theory of relativity is applicable and that the internal motions of the star are non-relativistic. Furthermore, the long-wavelength approximation was employed, which in practice has to be verified case by case. Finally, we adopted the isotropic elastic sphere model of a star according to the general idea of combining the physics of spherical resonant mass gravitational wave detectors with asteroseismology. This approximation was made for calculative simplicity and to demonstrate the formalism of the proposed excitation mechanism.

The equation of motion for the internal velocity field was solved analytically and the calculations somewhat parallel the methods used in gravitational wave detector physics and those in earlier works on excitation mechanisms (\citealt{Balmforth1992}; \citealt{Goldreich1994}; \citealt{Samadi2001a}). It is found that external gravitational waves solely excite quadrupolar spheroidal normal modes and that eigenmodes are sensitive only to the spherical component $h_m$ of the gravitational waves having the same azimuthal order $m$. The main result of this paper is the expression for the stationary square oscillation amplitudes given by equations (\ref{eq:A2_f1}) and (\ref{eq:A2_f2}) and was derived under the assumption of a stationary external gravitational wave field. The square amplitudes were shown to have a characteristic $R^8$-dependence on the radius of the star (effective $R^2$-dependence), which endows this excitation mechanism with increasing importance for stars larger than the Sun, especially for those that are nearby strong gravitational wave sources. Furthermore, it is worth noting the remarkable simplicity of the main result: stationary square amplitudes are obtained by simply multiplying a factor that depends on the resonance properties of the star and that has an effective $R^2$-dependene with the power spectral density of the spherical accelerations of the gravitational waves incident on the star. The latter factor can be seen as an explicit manifestation of the excitations' tidal force nature.

Using the result for the square amplitudes, analytical expressions for the mean-square surface velocities were derived, which are analogous to the ones obtained by \citet{Samadi2001a}, \citet{Samadi2001b} and \citet{Belkacem2008,Belkacem2009} in case of stochastic excitation by turbulent convection. These surface velocities are crucial for assessing the astrophysical and asteroseismic relevance of the proposed excitation mechanism. It is important to note, however, that the mean-square surface amplitudes derived in Section 6 are still theoretical quantities. In order to predict surface amplitudes due to the present excitation mechanism by gravitational waves that can be compared to observational data, one has to consider disc-integrated apparent surface velocities, which take both geometrical and limb-darkening effects into account. Since our calculations in the present paper are based on the elastic sphere model of a star, we do not proceed as far as to give detailed analytical expressions for the apparent surface velocities, although all important results derived in this paper such as equations (\ref{eq:A2_f1}), (\ref{eq:A2_f2}), (\ref{eq:v_s_f1}) and (\ref{eq:v_s_f2}) also hold in case of a hydrodynamical computation. 

In future studies, we will concentrate on a hydrodynamical framework of this excitation mechanism; preliminary results show that the aforementioned relations derived in this paper remain valid. This would then allow us to explicitly calculate apparent surface velocities and thereby to assess the asteroseismic relevance of this excitation mechanism.

\section*{Acknowledgments}
The authors are grateful to Oskar von der L\"uhe for valuable discussions. MR acknowledges support from the European Helio- and Asteroseismology Network (HELAS), which was funded as Coordination Action under the European Commission's Framework Programme 6.

\appendix
\section[]{Detailed discussion of the spatial gravitational wave integral}
This appendix is concerned with evaluating the spatial integral in equation (\ref{eq:A_1}), which is of the type
\begin{equation}
	\ddot{h}_{ij}(t)\left[\int_V\rmn{d}^3x\, \rho(r)[\mathbf{\Psi}^*_N(\bmath{x})]^i x^j\right], \label{eq:integral}
\end{equation}
where $\ddot{h}_{ij}$ is in TT gauge. Clearly, there are two cases to be considered.

\subsection{Toroidal modes}
Inserting equation (\ref{eq:tor_modes}) into equation (\ref{eq:integral}) yields the expression
\[
	\ddot{h}_{ij}(t)\int_V\rmn{d}^3x\, \rho(r)c_{nl}(r) x_j\epsilon_{ipq}x_p\partial_q Y^*_{lm},
\]
where $\epsilon_{ipq}$ is the Levi-Civita symbol. Integrating by parts and using the symmetry properties of the metric tensor $\ddot{h}_{ij}$ and of the Levi-Civita symbol when contracting one obtains the result, that the expression given in equation (\ref{eq:integral}) vanishes in case of toroidal modes \citep[see][]{Mag}. Therefore, toroidal modes do not couple to external gravitational waves.

\subsection{Spheroidal modes}
For spheroidal modes one has to consider the expression
\[
	\ddot{h}_{ij}(t)\mskip-5mu\int_V\mskip-5mu\rmn{d}r\rmn{d}\Omega\,\rho(r)r\left[a_{nl}(r)x_ix_jY_{lm}^*-\rmn{i}b_{nl}(r)\epsilon_{ipq}x_px_j L_q Y_{lm}^* \right],
\]
which is obtained by inserting (\ref{eq:sphmodes}) into equation (\ref{eq:integral}). The solid angle is denoted by $\Omega$. In the following, we consider the contributions from the two terms in brackets separately; the expressions arising from the first and second term are denoted by I and II, respectively.

By making use of equation (\ref{eq:rel_2}), the first contribution (I) can be rewritten as
\[
	\ddot{h}_{ij}\mskip-5mu\int_V\mskip-5mu\rmn{d}r\rmn{d}\Omega\,\rho(r)r^3a_{nl}(r)\left(\frac{1}{3}\delta_{ij}Y_{lm}^*+\mskip-10mu\sum_{m'=-2}^{2}\mskip-10mu c_{ij}^{m'}Y_{2m'}Y_{lm}^*\right).
\]
In this expression the contribution arising from the first term in brackets vanishes, since with the help of equation (\ref{eq:tt_gauge}) one has $\ddot{h}_{ij}\delta_{ij}=\ddot{h}^i_{\phantom{i}i}=0$. The contribution arising from the second term is easily seen to be non-vanishing only for $l=2$ due to the orthogonality property of the spherical harmonics. Consequently, in order to obtain a simple analytical expression for I, we again start with the original expression but now setting $l=2$,
\[
	\mskip-5mu\int_V\mskip-5mu\rmn{d}r\rmn{d}\Omega\,\rho(r)ra_{n2}(r)\ddot{h}_{ij}x_ix_jY_{2m}^*,
\]
and substitute equation (\ref{eq:rel_1}) therein. Then, again using the orthogonality property of the spherical harmonics, one is left with
\[
	\rmn{I}=\ddot{h}_{m}(t)\int_{0}^{R}\rmn{d}r\,\rho(r)r^3a_{n2}(r).
\]
Finally we define a new variable, $z\equiv r/R$, such that the integral is independent of the stellar radius $R$. This yields
\[
	\rmn{I}=\ddot{h}_{m}(t)R^4\int_{0}^{1}\rmn{d}z\,\rho(z)z^3a_{n2}(z)\equiv\ddot{h}_{m}(t)R^4\chi_n^{(1)}.
\]

Proceeding as in the former case, we employ equation (\ref{eq:rel_2}) to gain the following expression for the second contribution (II):
\[
	\ddot{h}_{ij}\mskip-5mu\int_V\mskip-5mu\rmn{d}r\rmn{d}\Omega\,\rho(r)r^3b_{nl}(r)\epsilon_{ipq}\mskip-5mu\left(\frac{1}{3}\delta_{pj}+\mskip-10mu\sum_{m'=-2}^{2}\mskip-10mu c_{pj}^{m'}Y_{2m'}\mskip-5mu\right)\mskip-5mu\rmn{i}L_q^* Y_{lm}^*.
\]
Since $L_q Y_{lm}$ gives a linear combination of spherical harmonics with the same value of $l$, the contribution from the first term in brackets vanishes unless $l=0$ due to the orthogonality property of the spherical harmonics. However, for $l=0$ the coefficient function $b_{nl}(r)$ is zero, $b_{n0}(r)=0$ (see Section 4). Likewise, the contribution arising from the second term in brackets is seen to vanish unless $l=2$.

Knowing that II is non-vanishing only for $l=2$, we start again with the original expression,
\[
	-\ddot{h}_{ij}\int_V\rmn{d}r\rmn{d}\Omega\,\rho(r)rb_{n2}(r)\epsilon_{ipq}\epsilon_{qab}x_px_jx_a\partial_b Y_{2m}^*,
\]
where $L_q=-\rmn{i}\epsilon_{qab}x_a\partial_b$ has already been inserted. Making use of the cyclic property of the Levi-Civita symbol and contracting the epsilon tensors yields
\begin{equation}
	\ddot{h}_{ij}\int_V\rmn{d}r\rmn{d}\Omega\,\rho(r)rb_{n2}(r)\left[r^2x_j\partial_i Y_{2m}^*-x_ix_jx_p\partial_pY_{2m}^*\right],\label{eq:int_sph_1}
\end{equation}
where the second term in brackets vanishes, since $x_p\partial_p Y_{2m}^*=r\frac{\partial}{\partial r}Y_{2m}^*=0$. Substituting $\left(\mathcal{Y}_{pq}^{2m}\right)^*\partial_i\left(\frac{x_px_q}{r^2}\right)$ for $\partial_i Y_{2m}^*$ in equation (\ref{eq:int_sph_1}) (cf. equation (\ref{eq:sph.harm})), one obtains after some laborious algebra and again using equations (\ref{eq:expan.sph.harm.}), (\ref{eq:sph.harm}) and (\ref{eq:rel_1}): 
\begin{eqnarray}	
	\rmn{II}\mskip-10mu&=&\mskip-10mu-2\ddot{h}_mR^4\chi_n^{(2)}+2\mskip-10mu\sum_{m'=-2}^{2}\mskip-10mu \ddot{h}_{m'}\mathcal{Y}_{ij}^{2m'}\left(\mathcal{Y}_{ip}^{2m}\right)^*\nonumber\\ &&\times\int_{0}^{R}\rmn{d}r\,\rho(r)r^3b_{n2}(r)\int\rmn{d}\Omega\,n_jn_p,\nonumber
\end{eqnarray}
where $\chi_n^{(2)}$ is defined by
\[
	\chi_n^{(2)}\equiv\int_{0}^{1}\rmn{d}z\,\rho(z)z^3b_{n2}(z).
\]
With the help of the general identity
\[
	\int\rmn{d}\Omega\,n_in_j=\frac{4}{3}\pi\delta_{ij}
\]
and the orthogonality relation (\ref{eq:orth.rel}) the second contribution II can finally be written as
\[
	\rmn{II}=3\ddot{h}_m R^4\chi_n^{(2)}.
\]

Consequently, it was shown that the expression (\ref{eq:integral}) is non-vanishing only for spheroidal modes with $l=2$, in which case
\[
	\ddot{h}_{ij}(t)\left[\int_V\rmn{d}^3x\, \rho(r)[\mathbf{\Psi}^*_N(\bmath{x})]^i x^j\right]=\ddot{h}_m R^4\chi_n,
\]
where we defined
\[
	\chi_n\equiv\chi_n^{(1)}+3\chi_n^{(2)}=\int_{0}^{1}\rmn{d}z\,\rho(z)z^3\left[a_{n2}(z)+3b_{n2}(z)\right].
\]
Note that given a particular density profile $\rho(z)$ and coefficient functions $a_{n2}(z)$ and $b_{n2}(z)$, this quantity is independent of the stellar radius $R$.

\label{lastpage}

\end{document}